\documentclass[10pt]{article}

\usepackage{graphicx}
\usepackage{array}
\usepackage{booktabs}

\usepackage[ps2pdf, colorlinks, bookmarks, pdfview={FitH}, pdfstartview={FitH}]{hyperref}
\usepackage{amssymb}
\usepackage{amsmath}

\def\noi{\noindent}

\begin{document}

\markboth{G.A. Kohring}{Complex Dependencies in Large Software Systems}

\title{Complex Dependencies in Large Software Systems}

\author{G.A. Kohring\\
        NEC Laboratories Europe, NEC Europe Ltd.\\
         Rathausalle 10, D-53757 St. Augustin, Germany \\
        kohring@it.neclab.eu}
        %%gkohring@gmail.com}
\date{}
\maketitle

\maketitle

\begin{abstract}
%% Text of abstract
Two large, open source software systems are analyzed from the vantage point 
of complex adaptive systems theory.  For both systems, the full dependency
graphs are constructed and their properties are shown to be
consistent with the assumption of stochastic growth. 
In particular, the 
afferent links are distributed according to Zipf's law for both systems.
Using the Small-World criterion for directed graphs, it is shown that contrary
to claims in the literature, these software systems do not possess Small-World
properties. Furthermore, it is argued that the Small-World property is not of
any particular advantage in a standard layered architecture. 
Finally, it is suggested that the 
eigenvector centrality can play an important role in deciding which 
open source software packages to use in mission critical applications.
This comes about because knowing the absolute number of afferent 
links alone is insufficient to decide how important a package is
to the system as a whole, instead the importance of the linking package plays a
major role as well.

\end{abstract}

\bigskip
\noindent{\bf Keywords:}
Complex Adaptive Systems, Stochastic Growth, Small World Networks,
Software Engineering

\noindent{\bf PACS:}
89.75.Fb, 02.50.Ey, 89.20.Ff

\bigskip

\section{Introduction}
\label{sec:intro}

Scientist and engineers are increasingly turning to the metaphor of the
complex adaptive system (CAS) \cite[]{Doo96} in order to understand 
the behavior of very
large software systems involving large numbers of
components \cite{Mye03,Mou03,Whe03,Con07,Mai08,Nar08}.
A central tenet within the CAS paradigm is that the network of interactions 
between the constituent entities, together with their intrinsic properties, 
largely determines a system's emergent
behavior \cite{Bar99,Alb02,Chr07}. If the network of interactions is scale-free 
then it will combine the robustness
properties of a random network with respect to the loss of an arbitrary node,
together with the fragility properties of a binary network with respect 
to the loss of one of its hubs. If a network exhibits small-world
characteristics \cite{Wat98}, then information should flow efficiently at 
both global and local levels.

For software systems the network of interactions takes the
form of a dependency graph between the components. At the system level,
static dependency graphs play an important role in the build
and package management processes; hence understanding their properties 
has important practical consequences for system administration. Dependency
graphs also provide a finer grained view of the system than is usually 
obtained by looking at high level architectural diagrams.

In this paper,
the static dependency graphs for two relatively large software systems, 
Debian \cite{Deb08} and Maven \cite{Mav08} are examined.
Debian is an open source Linux distribution comprising over $22,000$ separate
software packages totaling more than 20 Gigabytes of source code. It is one of
the oldest Linux distributions and is still employed as a server OS, 
not to mention its immense popularity among Linux
enthusiasts.
Maven is, in the first place, a software project management tool, but it is 
also a collection of repositories for Java software. It is these repositories
which give Maven, as a project management tool, its strength, because they
offer all Java developers a convenient mechanism for tracking package
dependencies with minimal effort.  

Approaching these systems from the CAS viewpoint means attempting to
identify and understand their emergent properties.
Previous studies of the Debain distribution and collections of Java software
have concentrated upon the degree distributions of the afferent and efferent
links in the
dependency graph \cite{Val02,Mye03,Mou03,Con07,Mai08,Bax08}. All studies agree
that the afferent degree distribution follows a power law, but they obtain
significantly different exponents for Debian and Java software,
leading one to wonder if the differences are
due to some fundamental differences in how the software is constructed,
or if they simply 
reflect the small sample size used in previous studies of Java software.
The main motivation for using Maven is to have a well defined, standard set of
Java software, representing
a much larger collection of packages than used in previous studies. Like
the Debian distribution, the Maven repositories represent reproducible data
sets upon which other researchers can cross-check the results presented here.

The present work has two aims, first it examines the issue of growth models
for describing the distributions seen in the afferent and efferent degrees 
in an effort to determine whether or not
the different distributions seen in these quantities can be explained in terms
of a single model. Using completely separate models for describing the
distribution 
of afferent and efferent links as was the case in previous studies is not 
satisfactory as the links arise from within the
same software development process and should therefore be correlated.

The second aim of this paper is to look
beyond the degree distribution to examine small world and
centrality issues.  An agglomeration of all of these properties gives
indications regarding the stability, maintainability and maturity of
the various software packages.

\subsection{Previous Work}
\label{sec:previous}

Maillart \textit{et al.} \cite{Mai08} measured the afferent degree
distribution for 
the Debian system. Valverde \textit{et al.} \cite{Val02} examined Java
software and measured the total degree distribution by treating the dependency 
graph as an 
undirected graph.  Baxter \textit{et al.} \cite{Bax06,Bax08} and Wheeldon and 
Counsell \cite{Whe03} measured both the afferent and efferent degree
distribution for
different Java packages, whereby they made a more fine 
grained distinction between the different types of dependencies than is being
made in this paper.  Concas et al.\cite{Con07} also studied fine grained 
afferent and efferent degree distributions related to object-oriented software
quality metrics for Smalltalk software, while  
Myers \cite{Mye03} examined these quantities for several
open source software packages. In all of these studies, 
the afferent degree distribution
was found to obey a power law, while the efferent degree distribution,
the extent it was measured, was found to
follow either a power law \cite{Mye03,Whe03}, or some other distribution
\cite{Bax06,Bax08,Con07}. Even though a power law was
obtained for the afferent degree distribution
in all studies, there was no consensus on 
the exponent, with
various estimates ranging from a low of $1.4$ to a high of $2.5$.

Concas \textit{et al.}\cite{Con07}, Maillart \textit{et al.}\cite{Mai08}  and 
Baxter \textit{et al.} \cite{Bax08} proposed various growth
models to explain the origins of the observed phenomena.
Concas \textit{et al.} discussed random process involving both independent and
proportional growth models as well as the Yule type process \cite{Sim55}
suggested by Newman \cite{New05}.  They found
good agreement between the proportional growth model and the efferent degree
distribution,
as well as good agreement between the predictions of the Yule model and
the distribution of afferent degrees.
Maillart \textit{et al.} and Baxter \textit{et al.} also discuss Yule type
process, wherein the latter model is a discrete growth model while the 
former take a continuum approximation. The continuous model of 
Maillart \textit{et al.} was
applied to the afferent degree distribution
and showed good agreement with the
measurements.  The discrete mode of Baxter \textit{et al.} was applied to both
the afferent and efferent degrees
and while the agreement with measurements
for the afferent degree was quite good, the agreement with the measurements
for the efferent degree was not as satisfactory. 

The small-world properties have been previously measured for several open
source software packages 
by Moura et al. \cite{Mou03} and Valverde and Sol\'e \cite{Val03}. 
Both of these studies found evidence for small-world behaviour when
using the 
original Watts and Strogatz's \cite{Wat98} definition of a small-world
graph. To apply this definition, both studies first converted the directed
dependency graphs to undirected graphs.

To the best of our knowledge, eigenvector centrality has not previously 
been studied in the context of dependency graphs for large software systems.
Other methods for ranking software based upon multiple factors are discussed
by Tsatsaronis et al. \cite{Tsa09}.

\subsection{Scope of Present Work}

This paper addresses shortcomings in the previous work by demonstrating that
the afferent and efferent links in both the Maven and Debian software systems 
have the same distributions and that these distributions are shown to 
arise from the same simple growth model. Towards this end,
the next section discusses the construction of the dependency graphs for
both systems, while
Section \ref{sec:links}, discusses node centrality 
and the observed distribution of afferent and efferent links in more detail. 

Section \ref{sec:small} discusses the small world properties of
the two systems and shows why the small-world effects
seen in previous studies are an artifact of the conversion
from directed to undirected graphs.

Eigenvector centrality, another global
measure of centrality better known as the Google rank, is examined in 
Section~\ref{sec:google}.  The final section summarizes the results of 
this study discusses how the CAS
viewpoint can be applied to problems in software engineering.

\section{Constructing the Dependency Graph}
\label{sec:graph}

A dependency graph, $\bf G$ is defined as a pair of sets 
${\bf G}=({\cal N}, {\cal L})$,
where $\cal N$ is a set of nodes and $\cal L$ is a set of directed links.
If, $n, m\,\in {\cal N}$, then $(n,m)$ denotes a 
directed link from $n$ to $m$; i.e., a dependency of $n$ on $m$.
The
nodes of interest here are either packages in the Debian case, or
jar files (Java Archives) or classes in the Maven case.  The directed links 
symbolize the 
dependencies between the packages, jar files or classes. 
Define the set of afferent links for a given node $n$ as: 
${\cal D_A}(n)=\{ (m,n) | (m,n) \, \in \, {\cal L} \}$
and the set of efferent links as 
${\cal D_E}(n)=\{ (n,m) | (n,m) \, \in \, {\cal L} \}$.  
For convience, denote
the number of afferent links for $n$ by $q_{\cal A}(n) = |{\cal D_A}(n)|$ 
(or $q_{\cal A}$ when speaking of an arbitary node) and 
the number of efferent links by $q_{\cal E}(n) = |{\cal D_E}(n)|$. $n$'s
total degree is then $q(n) = q_{\cal A}(n) + q_{\cal E}(n)$.

There are three graphs of interest here: the Debian dependency graph, 
${\bf G}_D$, the Maven jars dependency graph, ${\bf G}_J$ and the 
Maven class dependency graph ${\bf G}_C$. Each of these graphs is constructed
from the binary packages without recourse to the source code.

Determining ${\bf G}_D$ is straight forward since the 
Debian package management system \cite{Deb08a} requires the dependency 
information in order to install the packages,
the information is readily available in the form of a control file
accompanying each package and collated into a single \textit{Package} file
for each directory of the repository. For each package, its control file list
the packages upon which it depends.  There are five types of dependencies
defined in the Debian policy manual:
\textit{Depends, Recommends, Suggests, Enhances, Pre-Depends}. For the present
purposes no distinction is made between the different types of dependencies,
rather they are all used to build the edges of the ${\bf G}_D$.

Within a Debian control file it is also possible to define so-called
\textit{virtual packages}, which are logically existing packages whose
functionality is provided by some concrete package.  Virtual packages are
defined through the \textit{Provides} field in the control file. For purposes
of this study, virtual packages are treated as concrete packages.

Determining ${\bf G}_J$ and ${\bf G}_C$ is more involved.  
The main repositories are housed
at maven.org, codehouse.org and javax.org. Together these repositories contain
some $22,000$ jar files from more than $4,000$ separate projects.
Although Maven defines a
project management file (POM) in XML format which can be used to construct 
${\bf G}_J$ in
a manner similar to how ${\bf G}_D$ was constructed, it is more instructive to 
first construct ${\bf G}_C$ directly, then use the information 
therein, combined with the POM file to construct ${\bf G}_J$. Towards this end, 
the jar files for each project
are opened and the binary class files are parsed to determine their 
dependencies on other classes. Again, in software engineering one would make
a distinction between those dependencies which appear on the class' public 
interface, those which are private to the class and those which are local to
a class method \cite{Bax06,Col07,Con07}. For the purposes of the present 
study the technical
differences between these different types of dependencies are ignored.

For readers interested in confirming the results presented here, the Debian 
distribution used in this study was version 5.0.0, also known as 
\textit{Lenny}. Its
first official release was on February 14, 2009. (Actually, this study was
started using a beta version of the \textit{Lenny} release, but the results
were reconfirmed following the first official release.) The Maven repositories
are in a more continual state of flux and the data used here was taken during
the first week of January 2009.

\section{Degree Centrality}
\label{sec:links}

The degree centrality of node $n$ is defined as the size of the set of links 
starting or ending at $n$. It plays an important role in the
theory of random networks \cite{Alb02}. 
(Throughout this paper the terms ``graph'' and ``network'' will be used
interchangeably.) 
When the distribution of  
$q(n)$ over all $n \in {\bf G}$ follows
a power law, then network's tolerance to random failures is increased. For
directed networks, the distribution of afferent and efferent needs to be
treated separately.

\subsection{Afferent Links}

The distribution of afferent links found in this study is depicted in 
Fig.~\ref{fig:debAff} 
for Debian and Fig.~\ref{fig:mavAff} for Maven. Throughout this paper the
complementary cumulative distribution function, $P(X>x)$, is
measured in lieu of the probability mass distribution, $p(x)$; as the former 
provides 
more accurate estimates of the distribution's parameters when the data is
noisy \cite{Gol04,New05,Cla09}. The first point to notice is that for 
all ${\bf G}$, $p(q_{\cal A})\sim
q_{\cal A}^{-\alpha}$ with $\alpha\approx 2$ ($P(Q_{\cal A} > q_{\cal A})\sim
q_{\cal A}^{-\alpha + 1}$). Such a distribution
generally goes by the name of Zipf's law and has been previously found to hold
for a number of different natural and anthropological 
phenomena \cite{New05}. 
In fact Maillart \textit{et al.} \cite{Mai08} have previously demonstrated 
that this 
relationship holds for the Debian distribution.

What is new in these results is that Zipf's law holds for Maven as 
well.  Using the MLE (maximum likelihood estimate)\cite{Gol04,Cla09}, an
accurate estimate of $\alpha$ can be obtained yielding $\alpha=2.0\pm 0.1$,
which holds for both the Debian and Maven systems.
Previous studies on a number of different Java software have found
exponents ranging from a low of $\alpha\approx 1.4$ to a high of
$\alpha\approx 2.5$ \cite{Val02,Mye03,Whe03,Bax06,Con07}.  The discrepancy
between past results and the current experiments are
due to two factors: 1) previous studies limited the types of dependencies 
they investigated, and 2) previous studies examined only single
Java projects, not the broad collection available under Maven. Limiting the
dependency counting to those on the public interfaces suppresses the
distribution at large values of $q_{\cal A}$ since the majority
of dependencies are hidden as local variables inside the body of 
class' methods.  Finding the same
exponent for Maven and Debian is reassuring as it indicates that similar
mechanisms are involved in the engineering processes through which the
dependency graphs arise.

\subsection{A Yule Process}

The exact mechanisms for generating Zipf's law behavior are still under
debate\cite{New05,Ree03,Mit04}.  Some suggestions include various 
Yule processes \cite{Sim55,Pri76}, entropy maximization \cite{Har01} and 
highly optimized tolerance \cite{Doy00}. 
For Graphs, Barab\'asi and Albert \cite{Bar99} defined a variant of Simon's
Yule process \cite{Sim55}, called \emph{preferential attachment}, which leads 
to power law behavior in the distribution of the
afferent and efferent links, and Bollob\'{a}s \textit{et al.} \cite{Bol03}
extended their model to directed graphs. Although preferential attachment is
the most popular model, other discrete growth models similarly lead to power
law behavior in the afferent link distribution \cite{Kra01,Tad01}.
On the other hand,
Maillart \textit{et al.} \cite{Mai08} previously demonstrated that 
a continuous formulation
of the Yule process offered a good explanation
for describing the distribution $p(q_{\cal A})$.
Here, the aim is to provide further support for the latter hypothesis
by showing that it can explain the distribution of $p(q_{\cal E})$ as well.

The essence of all Yule processes is the Gibrat principle \cite{New05,Sai09}: 
the probability that an entity will experience an increase in the value of 
one its properties in the
next time step is proportional to the current value of that property. In other
words, \textit{the rich get richer}.
This principle leads to a stochastic formulation for the growth,
the details of which vary with the particular
model.  To start with, let $X=q_{\cal A}$, then the growth in $X$
varies over time as according to the stochastic equation:

\begin{equation}
dX=\mu X\,dt + \sigma X\, dW_t,
\label{eq:gibrat}
\end{equation}

\noi where $\mu$ and $\sigma$ are constants and 
$W_t$ is a standard
Wiener process, meaning $W_0=0$, $W_t$ is continuous, and
$W_{t+\tau}-W_t \sim N(0,\tau), \, \forall \, t,\tau >0$.
Although $X$ is in the present case discrete,
eq.~\ref{eq:gibrat} is nevertheless a good approximation when $X$ is large.
With the help
of the Ito Lemma, eq.~\ref{eq:gibrat} can be readily integrated to obtain the
probability mass distribution:

\begin{equation}
p(x;T)=\frac{1}{x\sqrt{2\pi\sigma^2 T}} \, 
        e^{-\frac{[\ln{(x)}+(1-2\mu/\sigma^2)\sigma^2T/2]^2}{2\sigma^2 T}},
\label{eq:logNorm}
\end{equation}

\noi where $T$ is the time period over which the stochastic growth occurs and
the initial value of $X(t)$ is taken to be $X(0)=1$. 

Note that individual sets of afferent
links, ${\cal D_A}(n)$, have disparate histories, i.e., they have been growing 
for different lengths of time, with
the oldest packages more than 15 years old and the youngest not more 
than a year; hence, to compare with the data, one needs to average over the
distribution of $T$ \cite{Bar99,Ada00,Sai09}.
Although it is possible in principle to track down the lifetime of each
package, and thereby obtain the true distribution of $T$, the effort to do 
so would be prodigious. Instead one can make progress by assuming
all possible values of $T$ up to some large value, $T=\cal T$, which is the 
lifetime of the software systems themselves, are equally likely:

\begin{equation}
\begin{split}
p(x)&=\int_0^{\cal T} \frac{1}{\cal T}\, dT\, p(x;T) \\
& \qquad \\
&\approx \frac{4\sigma^2}{{\cal T}(1-\frac{2\mu}{\sigma^2})}\,
\begin{cases}
      x^{-1} & \text{if $x<1$,}\\
      x^{-2 +\frac{2\mu}{\sigma^2}} &
                \text{if $x\ge 1$},
\end{cases} \\
& \qquad \\
& \qquad \qquad 
  - O\left( e^{-2\sigma^2\left[1-\frac{2\mu}{\sigma^2}\right]^2{\cal T}}\right)
\\
\end{split}
\label{eq:powx}
\end{equation}

\noi Eq.~\ref{eq:powx} holds for $\mu<\sigma^2/2$ and large $\cal T$. Under
the latter condition, the first term in eq.~\ref{eq:powx} dominates and all 
higher order terms in $\cal T$ can be neglected.

When $\mu \ll \sigma^2/2$, the growth is
dominated by the stochastic fluctuations and $p(x)\sim x^{-2}$, in 
good agreement with the measurements.

\subsection{Efferent Links}

The distribution of efferent links, $p(q_{\cal E})$, is depicted in 
Fig.~\ref{fig:debEff} for Debian and Fig.~\ref{fig:mavEff} for Maven.
Figs.~\ref{fig:debAff}-\ref{fig:mavEff} demonstrate clearly that the
distribution of the number of afferent links is not commensurate with the
distribution of the number of efferent links. The number of efferent links 
follows a lognormal distribution in agreement with \cite{Con07,Bax08}.

At first glance this might cast doubt on the explanation of stochastic growth
presented in the previous section; however, an examination of the assumptions
leading from
eq.~\ref{eq:logNorm} to eq.~\ref{eq:powx} reveals a difference between the
two. The set of afferent links, ${\cal D_A}(n)$, for a given node, $n$, grows
when another node, $m$, links to it. In software engineering, this linkage
occurs if the package or class represented by $n$ has services required by
the package or class represented by $m$.  Over time, as the size of Debian
or Maven collection increases, ${\cal D_A}(n)$ will continue to grow as
new packages are built using the services of existing packages. 
The set of
efferent links, ${\cal D_E}(n)$, on the other hand, expands only when the
\textit{responsibilities} of package $n$ expand. In software engineering, it is
considered poor practice to extensively change a class's responsibilities 
once it has become well established in the community.  The preferred approach
is to
create a new class which extends the older class. This ``open to extension,
closed to change'' philosophy \cite{Joh04} implies that the 
set ${\cal D_E}(n)$  will grow
only for the relatively short time need for $n$ to become mature and widely 
accepted.
Assuming that this settling time, to be denoted by, $T_s$, is
constant, independent of
any particular node, and relatively short compared with time the newer
nodes have been in existence, then the distribution, $p(q_{\cal E})$, 
would be expect to have the form given in eq.~\ref{eq:logNorm} with 
$T_s$ replacing $T$.

Other models for the distribution of links in a dependency graph, such as 
some form of preferential attachment \cite{Kra01,Tad01,Bol03}, highly optimized
tolerance \cite{Doy00}, local optimization \cite{Val02} and entropy 
maximization \cite{Har01} predict the same
distribution for both
the afferent and efferent links. Hence, at least in the field of software
engineering, the model presented here appears to be the most suitable in the
sense that it is able to accommodate different distributions in the 
afferent and efferent links starting from a common mechanism for their growth.

\section{Small Worlds}
\label{sec:small}

${\bf G}_D$, ${\bf G}_J$ and ${\bf G}_C$ are all sparse graphs. If they 
were completely random,
then one would expect to observe a low degree of local clustering coupled
with a small, average path length between nodes. 
A so-called small world graph \cite{Wat98} on the
other hand, combines a \textit{high} degree of local clustering with a 
small, average path length between nodes.  Watts and Strogatz's original 
definition of a
small world network considered only connected graphs, with undirected links;
whereas all the graphs studied here contain directed links.  Furthermore, the
graphs are disconnected in the sense that is not possible to start at
an arbitrary node and visit any other arbitrary node while transversing
the links in their proper direction. (If the links were undirected, then 
${\bf G}_J$ and ${\bf G}_C$ would be connected. Directed graphs fulfilling
this condition are often referred to as \textit{weakly connected}.)

An adequate definition for a small world signature in the case of directed 
graphs was given
by Latora and Marchiori \cite{Lat01}. In their paper they first defined 
the efficiency of a graph as:

\begin{equation}
E({\bf G})=\frac{1}{|{\cal N}|(|{\cal N}|-1)} \sum_{(n,m) \in{\cal L}} \, 
\frac{1}{d_{nm}}
\label{eq:globalEff}
\end{equation}

\noi where $d_{nm}$ is the directed distance from node $n$ to node $m$ and
$d_{nm}\ne d_{mn}$. As links in the networks studied here are not weighted,
$d_{nm}$ is simply the total number of links transversed in their
proper direction while walking form $n$ to $m$.
If there is no directed path
from $n$ to $m$, then by convention $d_{nm}=\infty$. (Note that ${\cal G}$ 
is directed acyclic graph, meaning it does not contain any circular 
dependencies, including self-dependencies.)
$E({\bf G})$ 
is normalized, $E({\bf G}) \in [0,1]$, with values near $1$ indicate highly
efficient information flow, while values near $0$ indicate information spreads
slowly through the network.

Having defined $E({\bf G})$ to measure how efficiently information flows 
through 
the entire graph, a similar quantity to measure the efficiency with which 
information flows locally within subgraphs can be
defined as $E({\bf G}(n))$, where ${\bf G}(n)$ is the subgraph of ${\bf G}$
consisting of all the nearest neighbors of $n$ together with their 
mutual links, but not $n$ itself, nor any links to $n$.

Small world graphs are then defined as those having large values of 
both global, $E({\bf G})$, and local efficiency, $\langle E({\bf G}(n)) \rangle$.
A random graph on the other hand would be expected
to exhibit relatively large values of $E({\bf G})$ but relatively small values
of $\langle E({\bf G}(n))\rangle$; while 
a well ordered graph should exhibit small values of $E({\bf G})$ and large
values of $\langle E({\bf G}(n))\rangle$. Note: just as in the original
Watts and Strogatz definition, the Latora and Marchiori definition does not
provide exact definitions for ``small'' and ``large''. Generally, their paper 
consider values less than $0.1$ as ``small'' and those greater than
than $0.25$ as ``large''.

Table~\ref{tab:sw} lists the values of the local and global efficiency found
for the dependency graphs, ${\bf G}_D$, ${\bf G}_J$ and ${\bf G}_C$. As can be
seen, all of the graphs have large local efficiency, combined with small
global efficiencies; hence, these graphs do not fulfil the small world
conditions.  This result contrasts with previous studies claiming to have
uncovered the small world signature in software dependency 
graphs.\cite{Mou03,Val03}

The problem
arises in that previous studies 
relied on the original Watts-Strogatz definition of a small world in
terms of undirected graphs rather than the Latora-Marchiori definition for
directed graphs.  Simply treating directed links as undirected links gives 
a false impression of the efficiency with which information flows through
the network.

To further understand the previous point, calculate the Pearson 
correlation coefficient between
the number of afferent and efferent links at the end points of each link
in the network:

\begin{equation}
r_{\alpha \beta}({\bf G})= \frac{1}{|{\cal L}|-1}
\sum\limits_{(n,m)\, \in \, {\cal L}} 
\left( \frac{q_{\alpha}(n)- \langle q_{\alpha} \rangle}{\sigma_{q_{\alpha}}} 
\right)
\left( \frac{q_{\beta}(m) - \langle q_{\beta} \rangle}{\sigma_{q_{\beta}}} 
\right)
\label{eq:assort}
\end{equation}

\noi where $\sigma_{q_{\alpha}}$ is the standard deviation of $q_{\alpha}$ and
$\alpha, \beta \, \in \, \{ {\cal A, E} \}$. Eq.~\ref{eq:assort} reduces to 
Newman's assortative mixing coefficient \cite{New02} when 
$q=q_{\cal A}+q_{\cal E}$.  $r({\bf G}) \in [-1,1]$ with $r({\bf G})=1$ 
indicating a perfect positive correlation (assortative mixing \cite{New02}),
i.e., nodes with a given value of $q(n)$ connect only to 
other nodes with
the same value $q(n)$, and $r({\bf G})=-1$ indicating perfect
negative correlation (disassortative mixing \cite{New02})
i.e., nodes with small $q(n)$ connect only to nodes with 
large $q(n)$ or vice versa. 

Table~\ref{tab:sw} lists the four mixing coefficients for the networks studied
here. The main trend to notice in the data is the uniformly, relative large 
values of $r_{\cal EA}({\bf G})$ and the uniformly, relative small values of
$r_{\cal AA}$ and $r_{\cal AE}$.  These results indicate that the directed
links have a disassortative preference with
nodes of low efferent degree connecting to nodes of
high afferent degree. In such networks information flow is inhibited since
nodes with high efferent degree, tend to have low efferent degree thus
inhibiting the efficient spread of information at the global level.

\section{Google Rank}
\label{sec:google}

The degree centrality discussed in section~\ref{sec:links} provides clues
about how important a given node is to the network in that it measures how
many other nodes depend upon it.
However, this is not the full picture, since a node 
with a small $q_{\cal A}(n)$ can be equally important to the network as a
whole if nodes with large $q_{\cal A}(n)$ depend upon it. To understand this
phenomena one needs to quantify the importance of a node in a graph based on the
importance of the nodes linking to it. This problem was famously solved 
by Brin and Page.\cite{Bri98}  Their solution rests on a variation of the 
eigenvector centrality measure, which is
defined in terms of the principal eigenvector of the adjacency matrix, $\bf A$.

The entries of the adjacency matrix in normalized form are 
defined as $A_{nm}=1/q_{\cal E}(n)$ if $(n,m) \in {\cal L}$ and $0$ 
otherwise.  In the present case some nodes have no
efferent links, while others have no afferent links, implying $\bf A$ is
singular; therefore there is no guarantee that all the components of the 
principal eigenvector will be non-negative, which is a necessary prerequisite
for using components of the eigenvector as a centrality measure.  To 
overcome the problem of non-existent efferent links
add to $\bf A$, the matrix $\bf B$, defined as: $B_{nm}=1/|{\cal N}|$ 
if
$q_{\cal E}(n)=0$, and $0$ otherwise. To overcome the problem of no afferent
links, add to $\bf A$ the matrix $\bf C$ defined as: $C_{nm}=1/|{\cal N}|$ 
for all
$n,m$. Since all the ${\bf G}({\cal N}, {\cal E})$ constructed here are sparse,
$1/|{\cal N}|<<1/q_{\cal E}(n)$, $\forall n$; thus, the contributions from
$\bf B$ and $\bf C$ to $\bf A$ are small and should not significantly distort 
the centrality measure.  The new matrix is: 

\begin{equation}
{\bf P}=\gamma\left({\bf A}+{\bf B}\right)+\left(1-\gamma\right) {\bf C},
\label{eq:adj}
\end{equation}

\medskip
\noi and the eigenvalue equation now reads:

\begin{equation}
{\bf R}={\bf P}^{\text{T}}\, {\bf R}
\label{eq:google}
\end{equation}

\noi In this form, eqs.~\ref{eq:adj} and~\ref{eq:google}, define
the Google page rank. In order to understand the meaning of
eq.~\ref{eq:google}, one can expand it: 

\begin{equation}
R_{n}=\frac{1-\gamma}{|{\cal N}|} + \sum_{(m,n)\,\in\, D_{\cal A}(n)} 
    \frac{\gamma R_m}{q_{\cal E}(m)} + \sum_{q_{\cal E}(m)=0} 
\frac{\gamma R_m}{|{\cal N}|},
\label{eq:googleExp}
\end{equation}

\noi showing that~\ref{eq:google} has the desired property of weighting 
the rank of each node by the rank of the nodes which link to it. The last term
in eq.~\ref{eq:googleExp} adds a small weight from all nodes with no
efferent links, while the first term gives a small weight to nodes with no
afferent links.
The properties of $\bf P$ are well known \cite{Hav03}:
it is a large, sparse, column stochastic matrix whose dominant eigenvalue is 
equal to $1$; the eigenvector corresponding to the dominant
eigenvalue has only non-negative elements; and the second largest eigenvalue 
is $\gamma$.
As long as $\gamma >> 1/|{\cal N}|$, the exact value does not
influence the results, however, if a power method is used
to solve for the principal eigenvector, the rate of convergence is 
equal to $\gamma$; meaning the power method will fail to converge as 
$\gamma \rightarrow 1$. In this study the value $\gamma=0.9$ is used.

Figs.~\ref{fig:allRank} depicts the average rank as a function of the number
of afferent links for the various graphs considered here. As can be seen there
is a clear trend, with large numbers of afferent links correlating with high
average rank. However, the average values glosses over significant details of 
the graph structure as can be seen by examining
Table~\ref{tab:rank}. The number third ranked Debian package has a mere 8
afferent links, while the 7th ranked package has $5,620$. And the number 
6th ranked
jar file in the Maven repositories has a only $2$ afferent links while the 
file ranked 7th has $1,092$. The reason a node with a very small number of 
afferent
links can be ranked so highly is that other, more highly rank nodes depend
upon it.

\section{Summary and Conclusions}
\label{sec:summary}

This study has demonstrated three points. Firstly, the distribution of afferent
and efferent links for dependency graphs of large software systems are similar
independent of the details of how the system is constructed.
For the afferent links, this distribution obeys Zipf's
Law. For the efferent links, the distribution is lognormal.  Both of these
distributions can be explained in terms of the same stochastic growth 
process, with the differences in the final form of the distribution
explainable in terms of
the different time scales over which the growth takes place. From these
results one can hypothesize, that the dependency graph of any sufficiently 
large software 
system, will, from a complexity viewpoint, have the same properties.

Secondly, this study has shown that the small world metaphor does not apply to
large software systems, because the global efficiency is too small.  This 
result contrasts with previous work which used the small world definition for
directed networks, but can be understood by 
considering about the high level architecture of the Debian system. 
From the software engineering point of view, the 
Debian distribution is well structured, employing a layered architecture with
the Linux operating system on the bottom, and Gnome/GTK+ (or KDE/Qt) 
applications on the top.  
In principle, developers should target their 
applications at a
particular layer, building upon services in the layer below,
while offering services to the layer above. In practice, applications in a
given layer will often use services from any or all of the lower lying layers, 
but never from a higher layer.
The long-range links typical of small-world networks would, in a layered
architecture, proceed from higher layers
down to the lower layers.  Without the
reverse couplings information flows upward in the stack, but not downward;
thus stifling the global efficiency and the classical small-world effects. It
would be an interesting extension to develop a signature for layered networks 
similar in spirit to the signature for small world networks.

Thirdly, the importance of a node in a dependency graph depends not only on
the number of afferent links, but also on the importance of the node from 
which the afferent link arises.
For example, the \textit{glibc-doc} package has only 8~afferent links; 
however, it is ranked 3rd amongst all Debian packages because
the number one ranked package, \textit{libc6}, links to it.

To further understand how CAS considerations can play a more active role
in software engineering consider the questions of 
stability, maintainability and maturity raised in the introduction. In the 
context of object oriented software development, an important software quality
metric is the instability, which is defined as the ratio of the
number of efferent links to the sum of efferent and afferent links for a given
object \cite{Mar03}.  Within the CAS paradigm, one would apply this metric 
on the system,
as a whole using the methodology for determining the link
distribution outlined above. In the notation used here, the instability of 
node $n$, would be written as: $I(n)=q_{\cal E}/(q_{\cal E}+q_{\cal A})$. 
This would lead to a measure of instability for
each open source package and provide important clues about the risk of
using any given package in one's own project. Large values of $I(n)$, indicate 
an instable package which is likely to change more often than a package with
a relatively smaller value of $I(n)$.

The maintainability and maturity of a software package is not just a 
question of its age or the number of previous releases, but also a question of
its acceptance. If it is not being used by other projects, then it will
sooner or later fade away.  Determining which packages are more 
likely to be maintained and updated over time, is a key factor in
deciding whether or not to use open source software in mission-critical 
settings.
Tsatsaronis et al. \cite{Tsa09} tackle this question by creating a
model for open source software repositories containing 19 common parameters 
for each
package. While these parameters are very good at judging the current health of
an open source project, they do not provide enough insight into 
question of whether or not a package will likely be maintained in the long run. 
Successful projects will always look good according to the criterion of 
Tsatsaronis et al. as long as
the project's originator is still running the project; however, what happens
when the project's originator, for whatever reason, is no longer able to
coordinate the project? How likely is is that the software will be maintained?
This results shown here suggests that the answer to this question, is to
expand Tsatsaronis et al. parameter list to include the 
Google Rank of the package in the network of all open-source software.
The more central a given package is to the system as a whole, the more likely it
will be that a talented developer will step forward to maintain a package 
once its originator has left.

Finally, as discussed above, the layered architectures examined here do not
exhibit small-world properties for good reasons; however, it is an open
question whether or not other architectural patterns might benefit from
small-world behavior. In particular when using a Service Oriented 
Architecture (SOA) to establish an ecosystem of services \cite{Qui07}, the 
small-world property may have advantages and may arise naturally. The
important concept to consider is how information should flow in the system and
whether information should flow as freely over long distance as it does 
over short distances.

\bigskip
\bigskip
\bigskip

\appendix{\large\bf Acknowledgements}
\bigskip

\noi The author would like to thank Dr. L. Lo Iacono for many useful discussions
related to this work and the anonymous referees for their constructive
criticisms.

\newpage

\bibliographystyle{elsarticle-num}
\label{sec:bibliography}
\bibliography{cs}

\newpage
\appendix{\large\bf Tables}
\label{sec:Tables}

\bigskip
\bigskip

\setlength{\extrarowheight}{4pt}
\begin{table}[htbp]
\begin{center}
\caption{The global and local efficiencies, $E({\bf G})$ and $ \langle E({\bf
G}(n)) \rangle $ respectively along with the Pearson coefficients for the
links, e.g., $r_{\cal EA}({\bf G})$ measures the correlation of 
$q_{\cal E}$ and $q_{\cal A}$ between connected nodes.}
\bigskip
\label{tab:sw}
\begin{tabular}{c>{\centering}m{1.5cm}>{\centering}m{1.5cm}>{\centering}m{1.5cm}>{\centering}m{1.5cm}>{\centering}m{1.5cm}c}
\toprule %
\  & $E({\bf G})$ & $ \langle E({\bf G}(n)) \rangle $ &
$r_{\cal AA}({\bf G})$ & $r_{\cal EA}({\bf G})$ & $r_{\cal AE}({\bf G})$ & $r_{\cal EE}({\bf G})$ \\ \toprule %
${\bf G}_D$ & 0.017  & 0.41 &  0.055   & -0.17  & -0.056   & 0.21\\ \midrule
${\bf G}_J$ & 0.0087 & 0.37 & -0.00043 & -0.10  &  0.00072 & -0.020\\ \midrule
${\bf G}_C$ & 0.024  & 0.33 & -0.021   & -0.28  & -0.033   & -0.095 \\ \bottomrule
\end{tabular}
\end{center}
\end{table}

\bigskip

\setlength{\extrarowheight}{4pt}
\begin{table}[htbp]
\begin{center}
\caption{\label{tab:rank} Rank and number of links.}
\bigskip
\begin{tabular}{cccc}
\toprule
& \multicolumn{3}{c}{$q_{\cal A}$} \\
\toprule
%%rank & \multicolumn{2}{c}{${\bf G}_D$} & \multicolumn{2}{c}{${\bf G}_J$} & 
rank & ${\bf G}_D$ & ${\bf G}_J$ & ${\bf G}_C$ \\
\midrule
 1 & 21,014      & 3,498          & 128,390  \\
   & libc6       & rt             & String\\ \midrule
 2 &  6,068      &    45          & 143,574  \\
   & libgcc1     & jce            & Object\\ \midrule
 3 &      8      &    12          & 30,141  \\
   & glibc-doc   & script-api     &  Throwable\\ \midrule
 4 &    200      &    83          &  26,426  \\
   & locales     & tools          &  Exception\\ \midrule
 5 &     12      &    88          &  13,741  \\
   & libc6-i686  & jsse           &  IllegalArgumentException\\ \midrule
 6 &     86      &     2          & 18,868  \\
   & gcc-4.3-base& charsets       & StringBuilder\\ \midrule
 7 &  1,606      & 1,001          & 40,783  \\
   &  debconf    & groovysoap-all &  Class\\ \midrule\midrule
27 &    208      &   123           &       4 \\ 
   & netbase     &   geronimo-spec-j2ee & AbstractStringBuilder\\ \midrule
28 &    136      &    81           & 2,408  \\
   & libgdbm3    &  org-openide-util     & Float\\
\bottomrule
\end{tabular}
\end{center}
\end{table}

\newpage
\appendix{\large\bf Figures}
\label{sec:Figures}
\bigskip

\begin{figure}[htbp]
\begin{center}
    \includegraphics[angle=90,height=17.5cm,keepaspectratio]{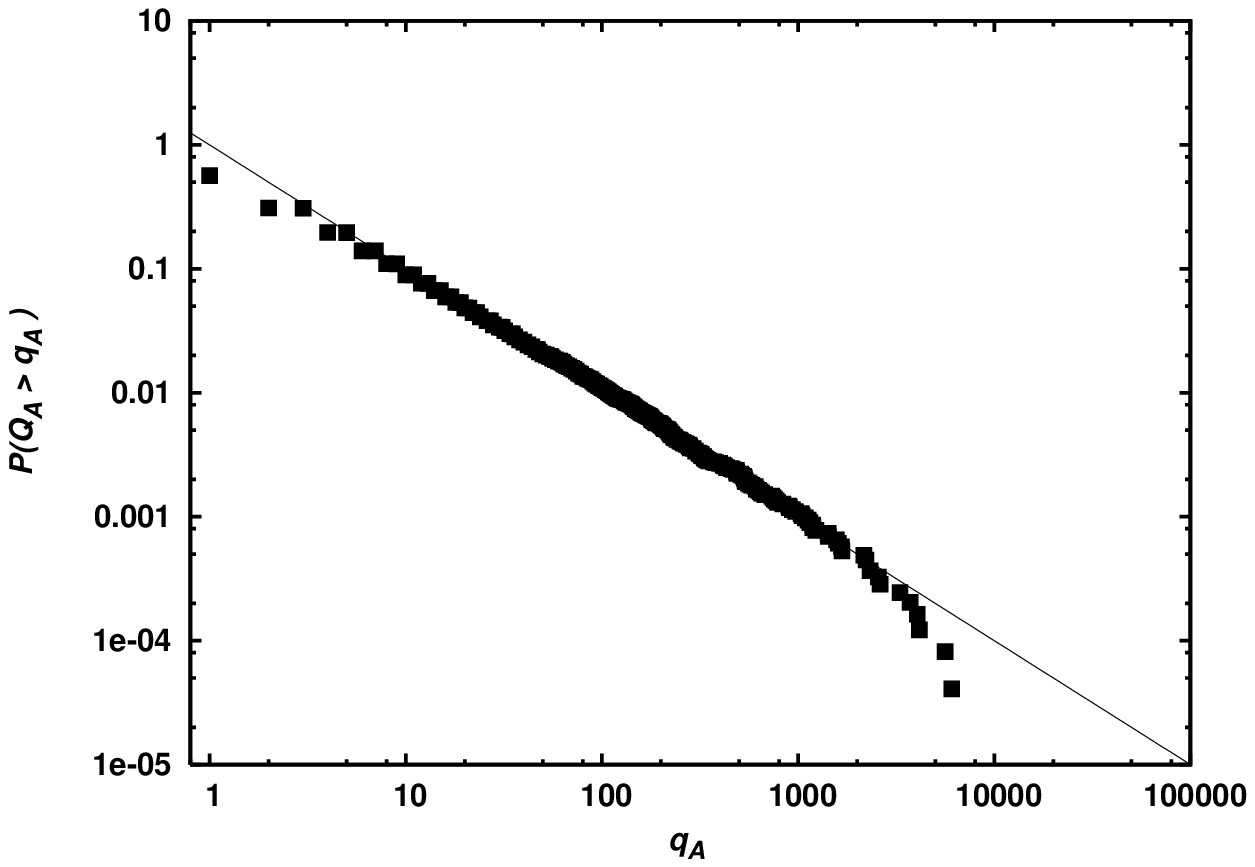}
    \caption{\label{fig:debAff}}
\end{center}
\end{figure}

\noi Figure~\ref{fig:debAff} Complementary cumulative distribution function,
$P(Q_{\cal A} > q_{\cal A} )$, for $q_{\cal A}$ in ${\bf G}_D$. The solid 
line, $P(Q_{\cal A} > q_{\cal A} ) = 1/q_{\cal A} $, is a guide to the eye.
\bigskip

\begin{figure}[htbp]
\begin{center}
    \includegraphics[angle=90,height=17.5cm,keepaspectratio]{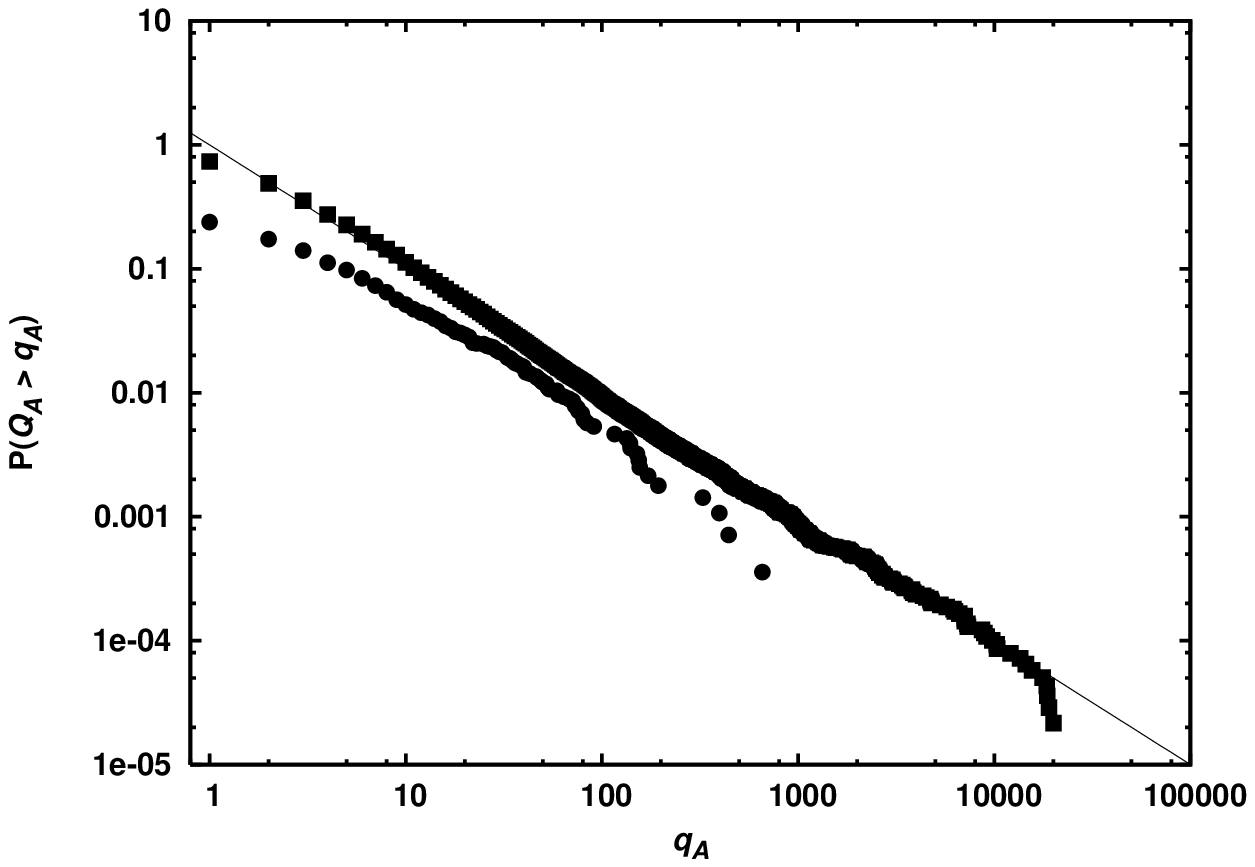}
    \caption{\label{fig:mavAff}}
\end{center}
\end{figure}

\noi Figure~\ref{fig:mavAff} Complementary cumulative distribution function,
$P(Q_{\cal A} > q_{\cal A} )$ for $q_{\cal A}$ in ${\bf G}_C$ (circles) and 
${\bf G}_J$ (squares).
The solid line, $P(Q_{\cal A} > q_{\cal A} ) = 1/q_{\cal A} $, is a guide
to the eye.
\bigskip

\begin{figure}[htbp]
\begin{center}
    \includegraphics[angle=90,height=17.5cm,keepaspectratio]{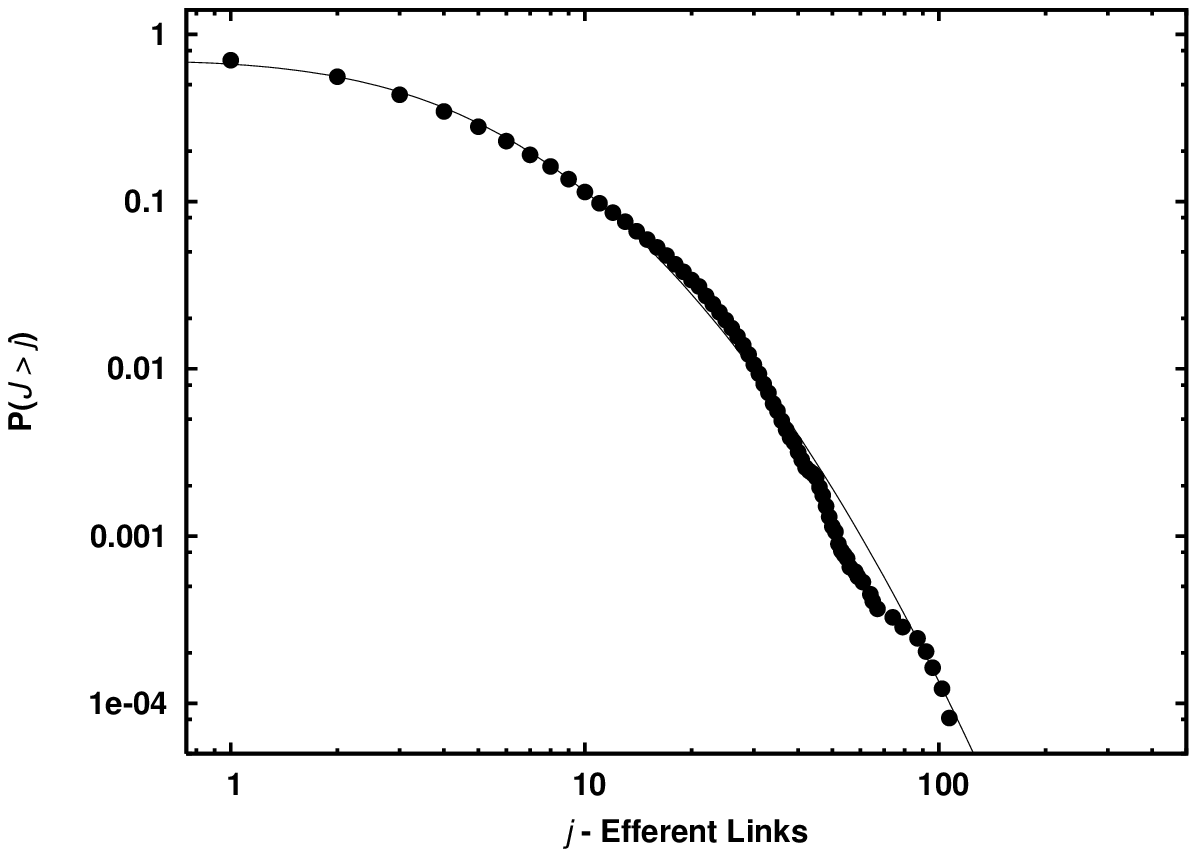}
    \caption{\label{fig:debEff}}
\end{center}
\end{figure}

\noi Figure~\ref{fig:debEff} Complementary cumulative distribution function,
$P(Q_{\cal E} > q_{\cal E} )$, for $q_{\cal E}$ in ${\bf G}_D$. The solid 
line is a best fit to $\text{erfc}\left( 
\frac{\ln(q_{\cal E}) - 2(1-2\mu/\sigma^2)\sigma^2T_s}{\sqrt{\sigma^2T_s}}
\right),$
which stems from eq.~\ref{eq:logNorm}.
\bigskip

\begin{figure}[htbp]
\begin{center}
    \includegraphics[angle=90,height=17.5cm,keepaspectratio]{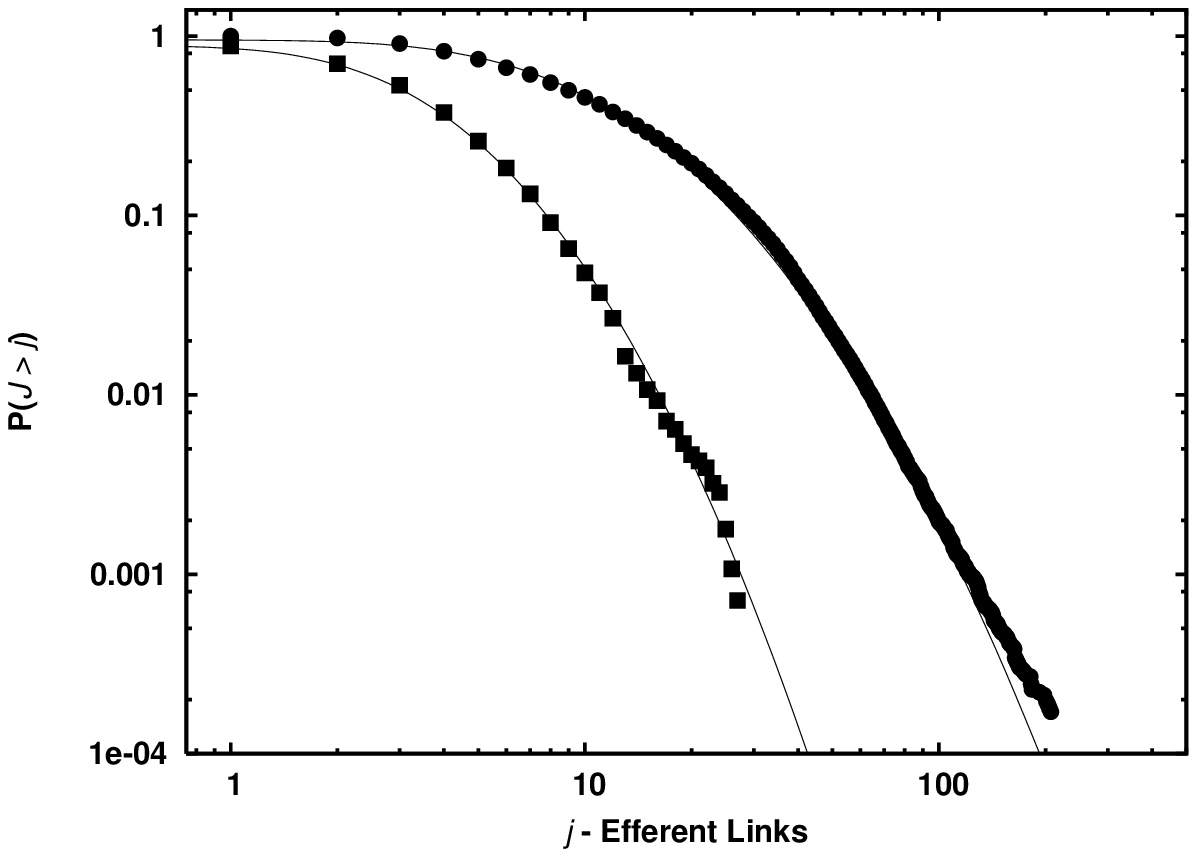}
    \caption{\label{fig:mavEff}}
\end{center}
\end{figure}

\noi Figure~\ref{fig:mavEff} Complementary cumulative distribution function,
$P(Q_{\cal E} > q_{\cal E} )$, for $q_{\cal E}$ in ${\bf G}_C$ (circles) and 
$\cal G_J$ (squares). 
The solid lines are best fits to $\text{erfc}\left( 
\frac{\ln(q_{\cal E}) - 2(1-2\mu/\sigma^2)\sigma^2T_s}{\sqrt{\sigma^2T_s}}
\right),$
which stems from eq.~\ref{eq:logNorm}.
\bigskip

\begin{figure}[htbp]
\begin{center}
    \includegraphics[angle=90,height=17.5cm,keepaspectratio]{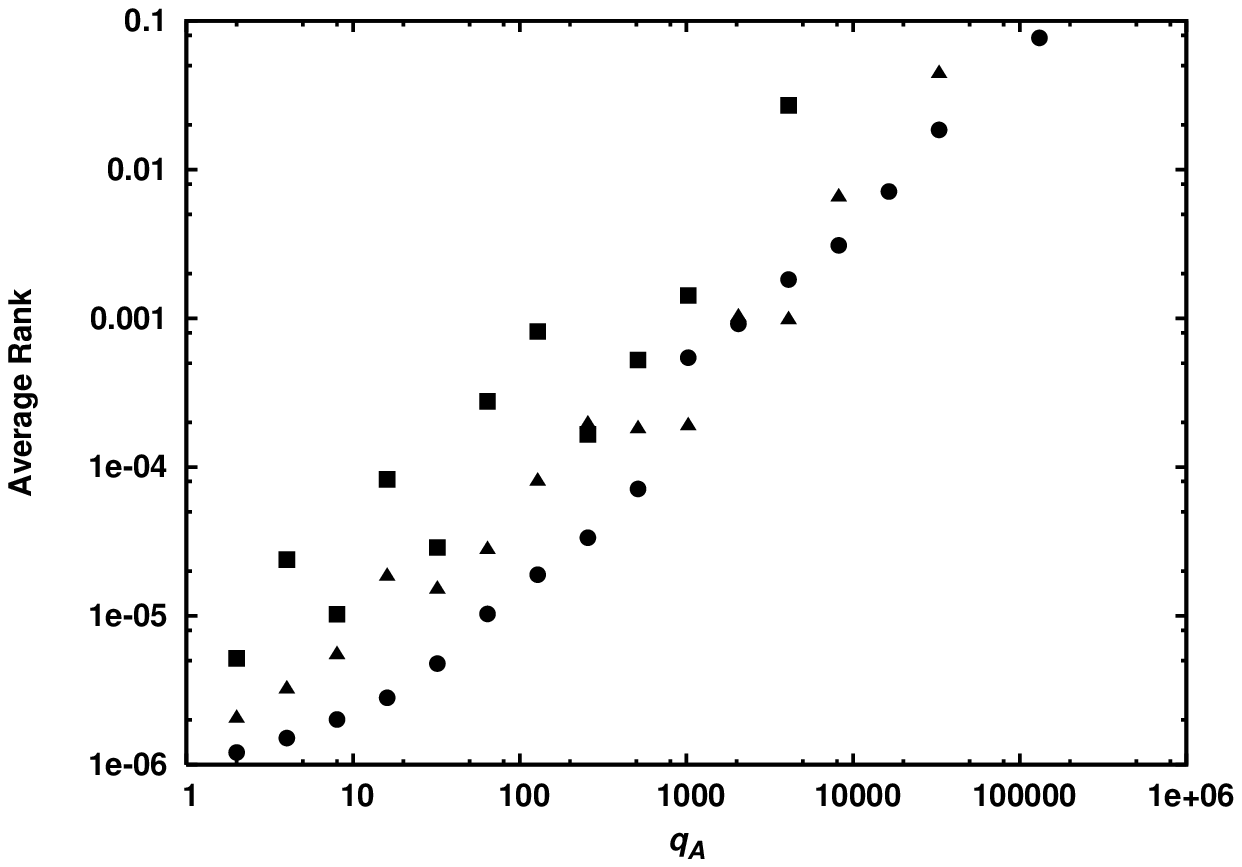}
    \caption{\label{fig:allRank}}
\end{center}
\end{figure}

\noi Figure~\ref{fig:allRank} Average rank versus $q_{\cal A}$. The circles
represent ${\bf G}_C$, the triangles, ${\bf G}_J$ and the squares ${\bf G}_D$.
\bigskip

\newpage

\newpage

\newpage

\newpage

\newpage

\end{document}